\documentclass[12pt]{iopart}
\usepackage{iopams}

\usepackage{revsymb}
\usepackage{graphicx}

\begin{document}

\title{Hot pixel contamination in the CMB correlation function?}

\author{R.~Aurich, S.~Lustig, and F.~Steiner}

\address{Institut f\"ur Theoretische Physik, Universit\"at Ulm,\\
Albert-Einstein-Allee 11, D-89069 Ulm, Germany}

\begin{abstract}
Recently, it was suggested that the map-making procedure,
which is applied to the time-ordered CMB data by the WMAP team,
might be flawed by hot pixels.
This could lead to a bias in the pixels
having an angular distance of about $141^\circ$ from hot pixels
due to the differential measuring process of the satellite WMAP.
Here, the bias is confirmed, and
the temperature two-point correlation function $C(\vartheta)$
is reevaluated by excluding the affected pixels.
It is shown that the most significant effect occurs in $C(\vartheta)$
at the largest angles near $\vartheta = 180^\circ$.
Furthermore, the corrected correlation function $C(\vartheta)$ is applied
to the cubic topology of the Universe, and it is found
that such a multi-connected universe matches the temperature correlation
better than the $\Lambda$CDM concordance model,
provided the cubic length scale is close to $L=4$ measured in
units of the Hubble length.
\end{abstract}



\section{Introduction}

One of the most important observational inputs of
modern cosmology is provided by
the cosmic microwave background (CMB).
The sky maps produced by the WMAP team
\cite{Hinshaw_et_al_2008} have found wide applications
in cosmological parameter extraction and provided
several surprises.
Besides the low power at large scales,
which had already been discovered by COBE \cite{Hinshaw_et_al_1996},
some strange properties emerged concerning the statistical isotropy,
i.\,e.\ an anomalous alignment between the quadrupole and the octopole
\cite{Tegmark_deOliveira_Costa_Hamilton_2003,%
deOliveira-Costa_Tegmark_Zaldarriaga_Hamilton_2004} as well as
an asymmetry of the CMB fluctuations within the
two hemispheres with respect to the ecliptic plane 
\cite{Eriksen_et_al_2004}.
Both anomalies are difficult to reconcile within the standard
inflationary framework.
A possible explanation could be related to the uncertainties
which inevitably arise in the foreground removal procedures
\cite{Gold_et_al_2008}.

Recently, Liu and Li \cite{Liu_Li_2008a,Liu_Li_2008b} have discussed a
possible systematics which lies at the heart of the map-making procedure.
The main point arises from the fact that WMAP measures
temperature differences at two points on the sphere separated
by about 141$^\circ$ which is due to the construction of the probe.
They demonstrate in \cite{Liu_Li_2008a,Liu_Li_2008b}
that the pixels,
which lie on a scan ring having diameter 141$^\circ$
with a hot pixel at its centre,
systematically obtain a lower temperature.
They show that the corresponding cross-correlation function
displays a negative correlation at the crucial angle of 141$^\circ$
in the three frequency bands Q, V, and W
thus pointing to a systematic distortion occurring in all three bands.
In order to estimate the magnitude of the induced temperature distortion,
they \cite{Liu_Li_2008a,Liu_Li_2008b} select the 2000 hottest pixels
for a detailed analysis and find a temperature distortion
in the range between $11\mu\hbox{K}$ and $14\mu\hbox{K}$
which in turn translates into a 10\% difference in the quadrupole and
the octopole moments.
This systematics is not remedied by a prolonged observation time.
It should be noted, however, that the WMAP team \cite{Hinshaw_et_al_2008}
computes the average temperature of pixels,
which do not belong to the set of hot pixels,
by omitting the hot pixels using the processing mask,
so that it is not clear at which point of the map-making pipeline
this systematics arises.

In this paper, we construct a modified mask
in order to check whether the suggested systematics really exists.
Furthermore, the two-point temperature correlation function $C(\vartheta)$
is investigated with respect to this systematics.
With the temperature correlation function,
which is based only on those pixels that are not strongly affected
by this systematics,
we carry out a test of a cubic topology of our Universe.


\section{A modified Liu-Li mask}

The hot pixel systematics does not affect all pixels on the
corresponding scan rings in the same way.
To each hot pixel belongs a well defined scan ring, of course,
but due to the large number of hot pixels
there is an equally large number of affected scan rings,
which intersect each other, in general.
This is due to the large size of the scan rings
having diameters between $135^\circ$ and $144^\circ$
depending on the channel
\cite{Hinshaw_et_al_2003b},
see Table \ref{Tab:Ring_Diameters} for the individual values.
Since the suggested systematics leads to a negative bias
on a given scan ring,
the bias should be more pronounced at the intersections of
several affected scan rings.
In order to test this effect, a modified Liu-Li mask is
constructed by the following procedure.

\begin{table*}
\begin{minipage}{140mm}
\hspace*{2.5cm}\begin{tabular}{|c|c|c|c|c|}
\hline
channel  & Q1, Q2        & V1, V2        & W1, W4        & W2, W3 \\
\hline
diameter & $144.3^\circ$ & $140.5^\circ$ & $139.8^\circ$ & $142.0^\circ$ \\
\hline
\end{tabular}
\caption{\label{Tab:Ring_Diameters}
The diameters of the scan rings in dependece on the channels
\cite{Hinshaw_et_al_2003b}.
}
\end{minipage}
\end{table*}

We start with the five-year Q1, Q2, V1, V2, W1, W2, W3, W4 channel sky maps
which are not foreground reduced and thus contain the hot pixels.
These maps as well as the KQ75 and KQ85 masks are provided by the WMAP team
\cite{Hinshaw_et_al_2008} on the LAMBDA home page lambda.gsfc.nasa.gov.
Then all pixels above a given temperature threshold
$T_{\hbox{\scriptsize thres}}$ are selected as hot pixels and
their corresponding scan rings with the diameters
given in Table \ref{Tab:Ring_Diameters} are computed.
In this way 8 maps are obtained whose pixel values contain the
number $N_{\hbox{\scriptsize inter}}$ of corresponding hot pixels.
From these 8 maps we compute 3 average maps,
one for each of the three bands Q, V, and W
with a mean $N_{\hbox{\scriptsize inter}}$.
If a given pixel obtains more than $N_{\hbox{\scriptsize inter}}$
intersections from affected scan rings,
it gets masked out.
This mask depends then on both thresholds $T_{\hbox{\scriptsize thres}}$
and $N_{\hbox{\scriptsize inter}}$.
In a last step all pixels within the KQ75 or KQ85 mask are rejected.
The remaining pixels should be those being less affected by foregrounds
and therefore containing the safest cosmological information.

\begin{figure}
\begin{center}
\vspace*{-30pt}
\begin{minipage}{11cm}
\hspace*{-25pt}\includegraphics[width=10.0cm]{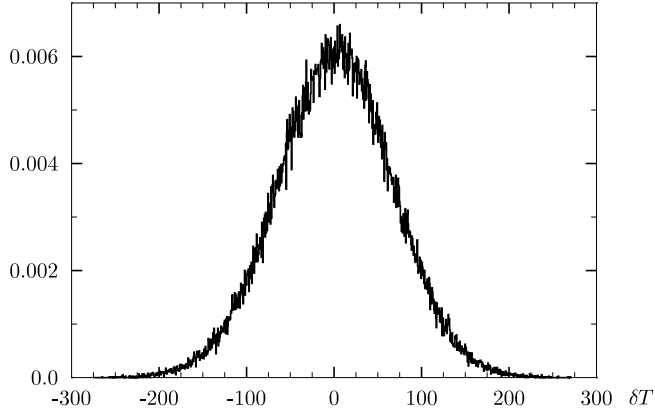}
\end{minipage}
\end{center}
\vspace*{-20pt}
\caption{\label{Fig:Histo_ILC_KQ75}
The distribution of the CMB temperature values $\delta T$ in $\mu$K
is shown for the ILC (5yr) map by taking into account only
the pixels outside the KQ75 mask.
The temperature distribution has a standard deviation of
$66\mu\hbox{K}$.
}
\end{figure}

\begin{figure}
\begin{center}
\vspace*{-5pt}
\begin{minipage}{9cm}
\hspace*{-10pt}\includegraphics[width=9.0cm]{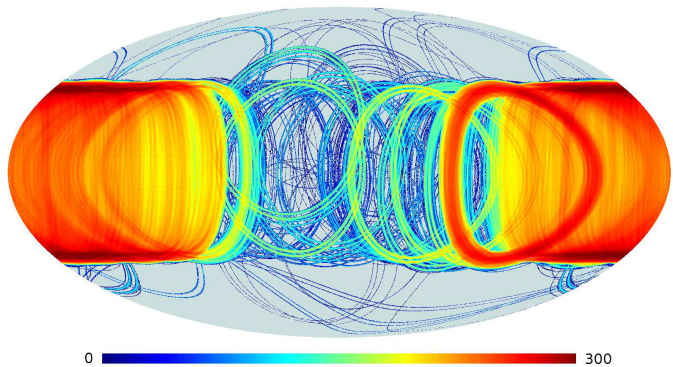}
\end{minipage}
\end{center}
\vspace*{-5pt}
\caption{\label{Fig:Liu_Lu_mask_1mk}
The degree of the disturbance by hot pixels is shown for the threshold
$T_{\hbox{\scriptsize thres}}=1\,000\,\mu\hbox{K}$ in Galactic coordinates.
The pixels are represented in dependence on the number
$N_{\hbox{\scriptsize inter}}$ of scan rings with respect to the W band
belonging to hot pixels with $\delta T>T_{\hbox{\scriptsize thres}}$
up to $N_{\hbox{\scriptsize inter}}=300$ (colour online).
}
\end{figure}

The value of the temperature threshold $T_{\hbox{\scriptsize thres}}$
is chosen to be many times larger than the typical temperature
fluctuation of the cosmological signal.
In Figure \ref{Fig:Histo_ILC_KQ75}
the distribution of the temperature values $\delta T$
is shown for the ILC map based on the five-year data
\cite{Hinshaw_et_al_2008}
by taking into account only the pixels outside the KQ75 mask.
The temperature distribution has a standard deviation of
$66\mu\hbox{K}$.
We choose in the following three thresholds of $1000\mu\hbox{K}$,
$2000\mu\hbox{K}$, and $4000\mu\hbox{K}$, respectively,
which are very large compared to the standard deviation of $66\mu\hbox{K}$
of the CMB signal.
In Figure \ref{Fig:Liu_Lu_mask_1mk}
we show the distribution of the affected pixels for the W band
for a hot pixel threshold
of $T_{\hbox{\scriptsize thres}}=1\,000\,\mu\hbox{K}$,
where the Mollweide projection is used in Galactic coordinates.
The intensity of the pixels depends on the number
$N_{\hbox{\scriptsize inter}}$ of hits by scan rings of hot pixels.
The scale is truncated at $N_{\hbox{\scriptsize inter}}=300$.
One observes that the most critical pixels lie below and above
the Galactic plane opposite to the Galactic centre
which is in the centre of Figure \ref{Fig:Liu_Lu_mask_1mk}.
This behaviour is caused by the large diameter of $\sim141^\circ$
of the scan rings and the fact that many hot pixels above
$T_{\hbox{\scriptsize thres}}$ are near the Galactic centre.
The modulus of Galactic latitude of the domain with the largest values of
$N_{\hbox{\scriptsize inter}}$ is given by $180^\circ-141^\circ=39^\circ$
revealed by the dark regions in Figure \ref{Fig:Liu_Lu_mask_1mk}.

\begin{figure}
\begin{center}
\vspace*{-10pt}
\begin{minipage}{11cm}
\hspace*{-25pt}\includegraphics[width=10.0cm]{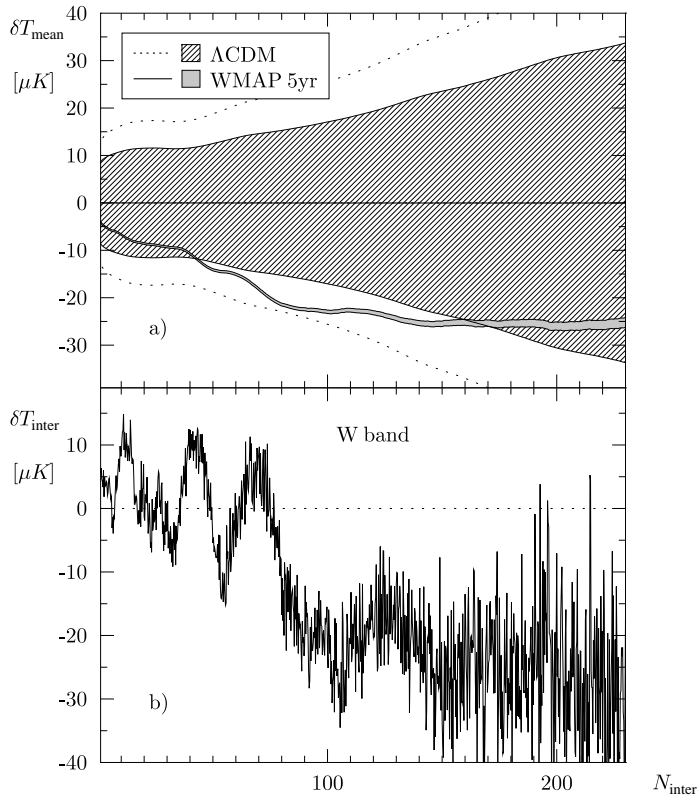}
\end{minipage}
\end{center}
\vspace*{-20pt}
\caption{\label{Fig:Mean_T_Liu_Li_Mask_W_band}
The mean temperature $\delta T_{\hbox{\scriptsize mean}}$ is shown
for the W band foreground-reduced map (5yr).
The mean value is computed from the pixels inside the modified Liu-Li mask
for the W band with $T_{\hbox{\scriptsize thres}}=1\,000\,\mu\hbox{K}$
but outside the KQ75 mask in dependence on the threshold
$N_{\hbox{\scriptsize inter}}$.
In Panel a) the value of $\delta T_{\hbox{\scriptsize mean}}$ is computed
from those pixels which are related to at least $N_{\hbox{\scriptsize inter}}$
hot pixels.
The grey band represents the $2\sigma$ uncertainty of the detector noise
in the corresponding area.
The shaded region denotes the 2$\sigma$ uncertainty obtained from
100\,000 simulations of $\Lambda$CDM models subjected to the same procedure.
The dotted curves show the 3$\sigma$ uncertainties.
In Panel b) the mean temperature $\delta T_{\hbox{\scriptsize inter}}$
is computed from pixels corresponding to exactly
$N_{\hbox{\scriptsize inter}}$ hot pixels.
}

\end{figure}

\begin{figure}
\begin{center}
\vspace*{-10pt}
\begin{minipage}{11cm}
\hspace*{-25pt}\includegraphics[width=10.0cm]{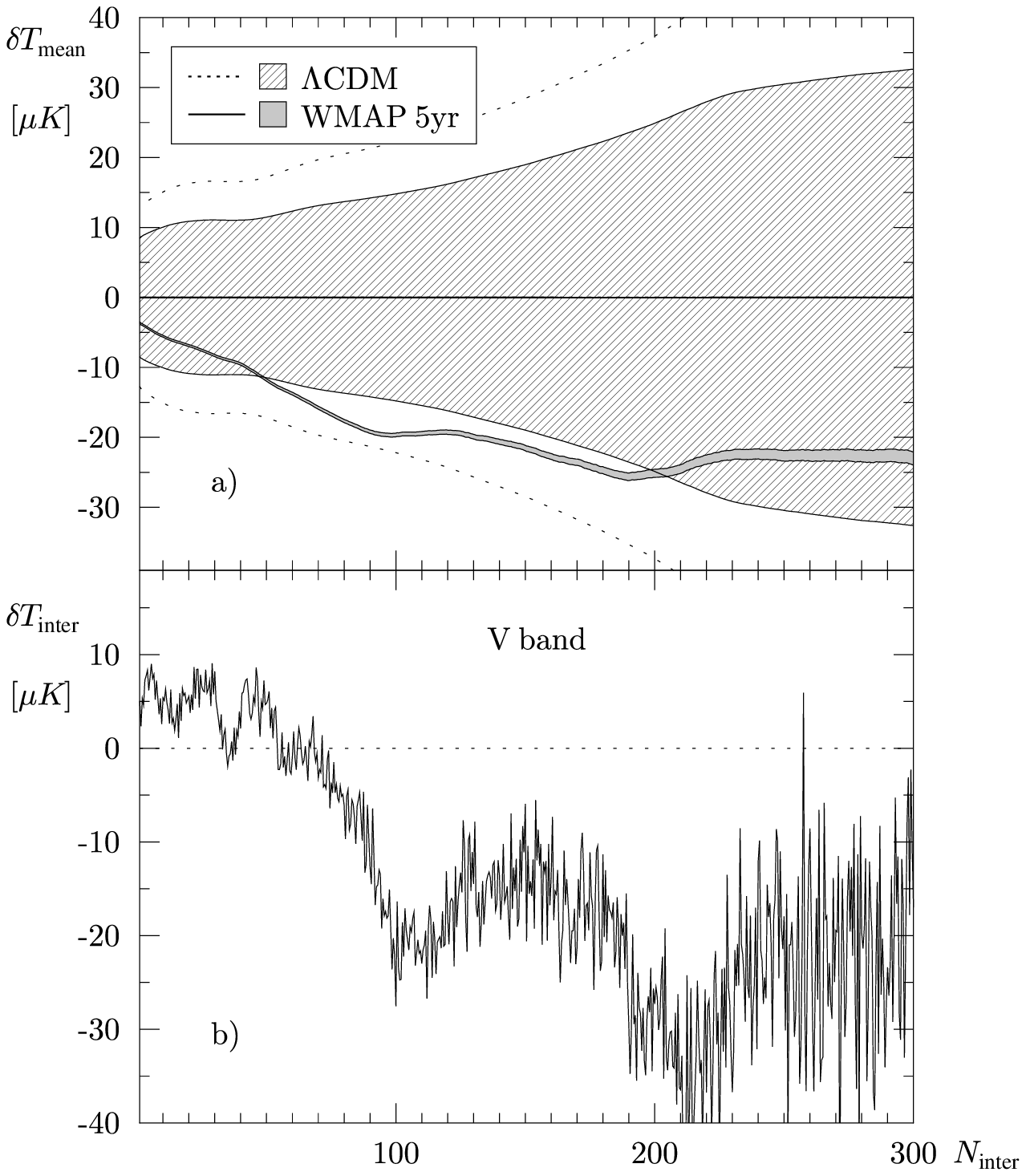}
\end{minipage}
\end{center}
\vspace*{-20pt}
\caption{\label{Fig:Mean_T_Liu_Li_Mask_V_band}
As  in Figure  \ref{Fig:Mean_T_Liu_Li_Mask_W_band}
the mean temperatures $\delta T_{\hbox{\scriptsize mean}}$
and $\delta T_{\hbox{\scriptsize inter}}$ are shown
but for the V band foreground-reduced map (5yr).
}
\end{figure}

\begin{figure}
\begin{center}
\vspace*{-10pt}
\begin{minipage}{11cm}
\hspace*{-25pt}\includegraphics[width=10.0cm]{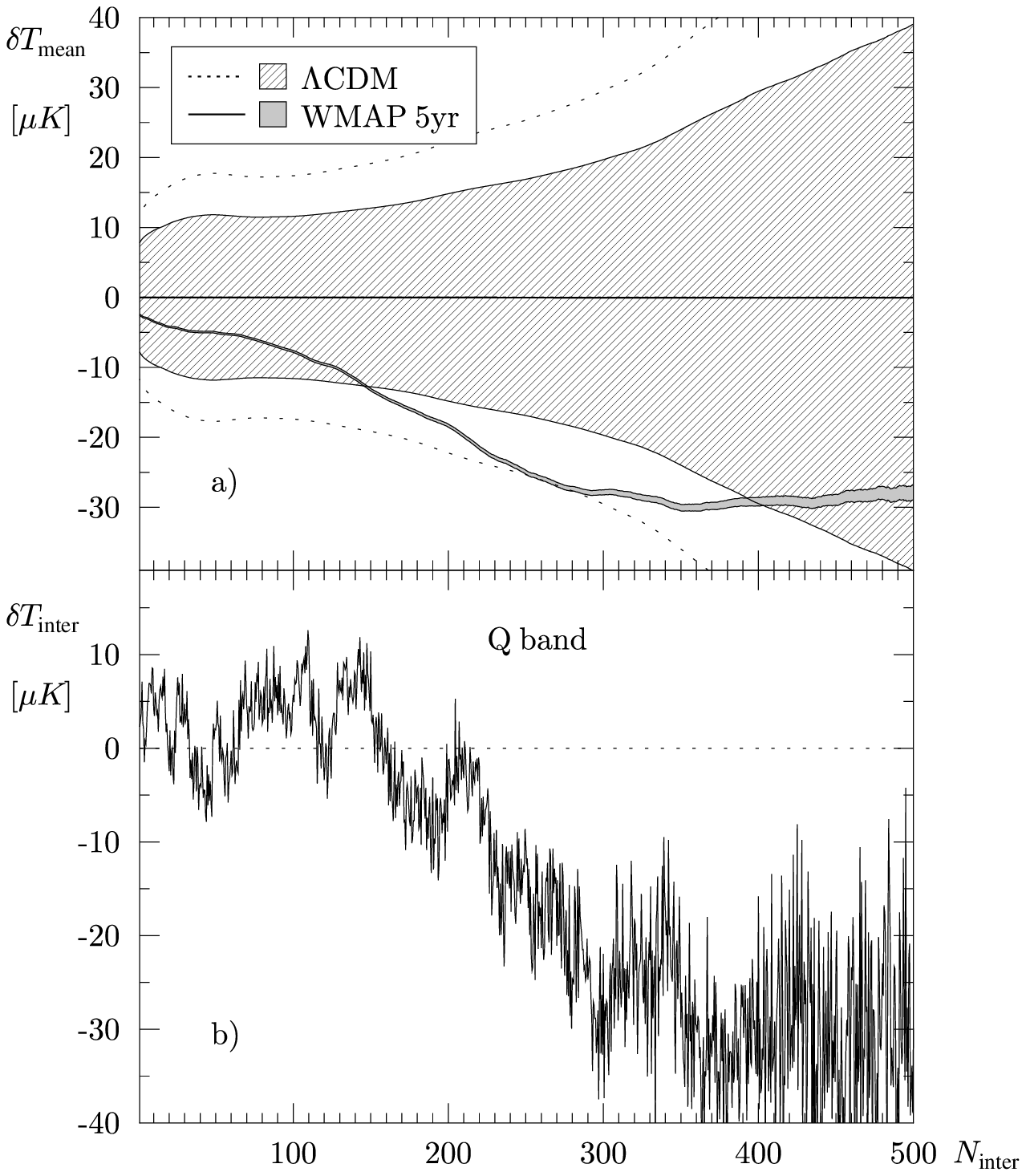}
\end{minipage}
\end{center}
\vspace*{-20pt}
\caption{\label{Fig:Mean_T_Liu_Li_Mask_Q_band}
As in Figure \ref{Fig:Mean_T_Liu_Li_Mask_W_band}
the mean temperatures $\delta T_{\hbox{\scriptsize mean}}$
and $\delta T_{\hbox{\scriptsize inter}}$ are shown
but for the Q band foreground-reduced map (5yr).
}
\end{figure}

In order to verify the negative-temperature bias of the pixels
belonging to hot pixel scan rings,
the mean temperature $\delta T_{\hbox{\scriptsize mean}}$ is computed
inside the modified Liu-Li mask with
$T_{\hbox{\scriptsize thres}}=1\,000\,\mu\hbox{K}$
but outside the KQ75 mask.
The mean $\delta T_{\hbox{\scriptsize mean}}$ depends on the chosen
value of the threshold number $N_{\hbox{\scriptsize inter}}$,
i.\,e.\ pixels are selected if they are related to at least
$N_{\hbox{\scriptsize inter}}$ hot pixels.
This procedure tests whether the negative bias increases
with the number $N_{\hbox{\scriptsize inter}}$ of scan rings.
Figure \ref{Fig:Mean_T_Liu_Li_Mask_W_band}a reveals
that $\delta T_{\hbox{\scriptsize mean}}$ computed
for the W band foreground-reduced map indeed drops towards more
negative values until a value of $N_{\hbox{\scriptsize inter}}\simeq 140$
is reached.
The pixels with $N_{\hbox{\scriptsize inter}}\gtrsim 140$
possess a temperature bias of the order of $-25\,\mu\hbox{K}$.
This is the expected behaviour
since with increasing $N_{\hbox{\scriptsize inter}}$
only the most severely affected pixels are taken into account,
i.\,e.\ those with the largest systematics.
Whereas panel a) displays $\delta T_{\hbox{\scriptsize mean}}$
computed from those pixels having {\it at least}
$N_{\hbox{\scriptsize inter}}$ hot pixels,
panel b) shows the mean value $\delta T_{\hbox{\scriptsize inter}}$
computed only from those pixels having {\it exactly}
$N_{\hbox{\scriptsize inter}}$ hot pixels.
The number of the latter ones is lower and thus
the curve possesses larger statistical fluctuations.
But it nevertheless demonstrates the negative bias for pixels
with $N_{\hbox{\scriptsize inter}}$ above 100.

In order to check the significance of this result,
100\,000 simulations of $\Lambda$CDM concordance models
are generated for which $\delta T_{\hbox{\scriptsize mean}}$
is computed using the same procedure as outlined above. 
These values scatter symmetrically around a zero mean.
The shaded region in Figure \ref{Fig:Mean_T_Liu_Li_Mask_W_band}a
denotes the 2$\sigma$ deviation from the zero mean.
One observes that the W band distortion is over a wide
range of values of $N_{\hbox{\scriptsize inter}}$ larger
than the 2$\sigma$ deviation for $\Lambda$CDM models.
This demonstrates that there is a systematics for these pixels.

Figures \ref{Fig:Mean_T_Liu_Li_Mask_V_band} and
\ref{Fig:Mean_T_Liu_Li_Mask_Q_band}
present the result of the same analysis as shown in
Figure \ref{Fig:Mean_T_Liu_Li_Mask_W_band},
but for the V and Q bands, respectively.
In both frequency bands the same negative bias of at least $2\sigma$
is observed when $N_{\hbox{\scriptsize inter}}$ is sufficiently large,
i.\,e.\ when enough hot pixels are connected to the considered one.
For very large values of $N_{\hbox{\scriptsize inter}}$
the deviation is smaller than $2\sigma$
but very few pixels occur in that range.
The values of $N_{\hbox{\scriptsize inter}}$ cannot be directly compared
between the different channels since they possess different hot pixels
lying above $T_{\hbox{\scriptsize thres}}$
because of the different foregrounds.
In the case of the W band the dust contribution dominates
whereas the synchrotron and free-free emission is more important
in the case of the V and Q bands.

None of our 100\,000 simulations of $\Lambda$CDM concordance models
yields a $\delta T_{\hbox{\scriptsize mean}}$ curve
which always has values below the WMAP result shown in the panel a)
of Figures \ref{Fig:Mean_T_Liu_Li_Mask_W_band} -
\ref{Fig:Mean_T_Liu_Li_Mask_Q_band}.
The same result is obtained for $\delta T_{\hbox{\scriptsize inter}}$
shown in the corresponding panel b).
This emphasises the peculiarity of the negative-temperature bias.


\section{The correlation hole at $\vartheta=180^\circ$}

Let us now come to the temperature two-point correlation function
$C(\vartheta)$, which is defined as
\begin{equation}
\label{Eq:C_theta}
C(\vartheta) \; := \; \left< \delta T(\hat n) \delta T(\hat n')\right>
\hspace{10pt} \hbox{with} \hspace{10pt}
\hat n \cdot \hat n' = \cos\vartheta
\hspace{10pt} .
\end{equation}
It was computed by \cite{Hinshaw_et_al_1996} from the COBE 4yr data
and was found to display unexpectedly low power at large angles
$\vartheta\gtrsim 60^\circ$.
At these large angles $C(\vartheta)$ is close to zero.
Only near $\vartheta=180^\circ$ it shows a slight negative power,
which is much more pronounced in the WMAP five-year data
\cite{Copi_Huterer_Schwarz_Starkman_2008}.
This anticorrelation of antipodal points is the so-called
``correlation hole''.
Copi et al.\ \cite{Copi_Huterer_Schwarz_Starkman_2008} show the strong
dependence of the correlation hole on the chosen mask.
The correlation function $C(\vartheta)$ possesses the largest
antipodal anticorrelation if one uses no mask at all,
whereas applying the KQ75 mask reduces the anticorrelation,
but it is nevertheless present.
They conclude that the full-sky results seem inconsistent with
the cut-sky results.
Furthermore, most of the large-angle correlation in
reconstructed sky maps lies inside the part of the sky
that is contaminated by the Galaxy.

We do not take into account the cosmic variance
which arises on the theoretical side since the cosmic initial
conditions are assumed only on statistical grounds.
Here, we are interested in the correlation function $C(\vartheta)$
observed from our special observer point within our Universe.

\begin{figure}
\begin{center}
\vspace*{-30pt}
\begin{minipage}{11cm}
\hspace*{-20pt}\includegraphics[width=10.0cm]{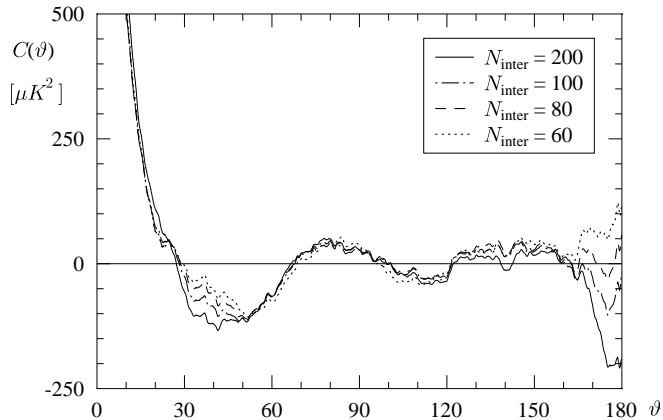}
\end{minipage}
\end{center}
\vspace*{-20pt}
\caption{\label{Fig:C_Theta_Liu_Li_Mask}
The correlation function $C(\vartheta)$ is computed from
the W band foreground-reduced map (5yr) using the pixels
outside the KQ75 mask.
In addition the pixels inside the modified Liu-Li mask are rejected
for 4 different values of the threshold $N_{\hbox{\scriptsize inter}}$.
}
\end{figure}

We now show that the remaining negative correlation near
$\vartheta=180^\circ$ is caused by the
pixels lying within the modified Liu-Li mask.
To that aim we apply our modified Liu-Li mask with
$T_{\hbox{\scriptsize thres}}=1\,000\,\mu\hbox{K}$
together with the KQ75 mask to the W band foreground-reduced map (5yr)
for several choices of the intersection threshold
$N_{\hbox{\scriptsize inter}}$.
The result shown in Figure \ref{Fig:C_Theta_Liu_Li_Mask}
reveals a pronounced correlation hole for large values of
$N_{\hbox{\scriptsize inter}}$,
i.\,e.\ for those cases
where only a small number of further pixels is rejected
after applying the KQ75 mask.
The curve belonging to $N_{\hbox{\scriptsize inter}}=200$
displays almost the usual result obtained by using only the KQ75 mask,
whereas the one belonging to $N_{\hbox{\scriptsize inter}}=60$ shows
even positive correlations near $\vartheta=180^\circ$.
The four cases presented in Figure \ref{Fig:C_Theta_Liu_Li_Mask}
also demonstrate that the correlation is at other angular separations
$\vartheta$ nearly independent on the value $N_{\hbox{\scriptsize inter}}$
except for the interval $[30^\circ,50^\circ]$.
However, the strongest sensitivity occurs at $\vartheta=180^\circ$.

Let us now address the question how probable such a behaviour
is with respect to the $\Lambda$CDM concordance model.
In order to compute the probability that $C(\vartheta)$ remains
nearly unchanged for $50^\circ<\vartheta<165^\circ$,
define the difference
\begin{equation}
\label{Eq:Delta_C}
\Delta C(\vartheta) \; := \;
C_{\hbox{\scriptsize KQ75+LL}}(\vartheta) \, - \,
C_{\hbox{\scriptsize KQ75}}(\vartheta)
\end{equation}
between $C(\vartheta)$ using the KQ75 mask and
the modified Liu-Li mask for  $N_{\hbox{\scriptsize inter}}=60$,
which is denoted as $C_{\hbox{\scriptsize KQ75+LL}}(\vartheta)$,
and $C(\vartheta)$ using only the KQ75 mask
denoted as $C_{\hbox{\scriptsize KQ75}}(\vartheta)$.
The integrated quantity
\begin{equation}
\label{Eq:Delta_C_int}
D \; := \; \int_{\vartheta=50^\circ}^{\vartheta=165^\circ} d\cos\vartheta \;
| \Delta C(\vartheta) |
\end{equation}
measures how strongly $C(\vartheta)$ is affected by the suspected pixels.
The value obtained from the W band data is surprisingly low,
indeed from our 100\,000 $\Lambda$CDM concordance models only 495 models
possess a smaller value,
i.\,e.\ $D_{\Lambda\hbox{\scriptsize CDM}}<D_{\hbox{\scriptsize WMAP}}$.
Thus, 99.5\% of the $\Lambda$CDM models show a stronger change in
$C(\vartheta)$ than the considered WMAP data.
Concerning the correlation hole, only 3111 models satisfy
$\Delta C_{\Lambda\hbox{\scriptsize CDM}}(180^\circ)>
\Delta C_{\hbox{\scriptsize WMAP}}(180^\circ)$,
i.\,e.\ the large increase of $C(\vartheta)$ at $\vartheta=180^\circ$
is also unusual at the 3 percent level.
If one requires both conditions simultaneously
not a single $\Lambda\hbox{CDM}$ model remains.
This emphasises the unusual behaviour of the suspected pixels.


\section{The cubic topology for the modified correlation function}

The surprisingly low CMB large-angle power of the
correlation function $C(\vartheta)$,
which was discovered in \cite{Hinshaw_et_al_1996},
is at variance with the $\Lambda$CDM concordance model
as has been found in \cite{Spergel_et_al_2003}
and recently emphasised in \cite{Aurich_Janzer_Lustig_Steiner_2007}
and \cite{Copi_Huterer_Schwarz_Starkman_2008}.
One of the possible explanations for this unexpected behaviour is
provided by models of the Universe whose spatial section is represented
by a multi-connected space form.
For a general introduction to the topology with respect to cosmology,
see e.\,g.\ \cite{Lachieze-Rey_Luminet_1995,Starkman_1998,%
Luminet_Roukema_1999,Levin_2002,Reboucas_Gomero_2004,Luminet_2008}.
In these models exists naturally a largest wavelength at which
perturbations can occur as seeds for structure formation.
This wavelength is mainly determined by the size of the topological
space form.
The correlation function $C(\vartheta)$ has been studied for several
multi-connected space forms, and a low large-angle power 
has been shown to exist in the case of
the hyperbolic Picard topology
\cite{Aurich_Lustig_Steiner_Then_2004a,Aurich_Lustig_Steiner_Then_2004b},
the spherical Poincar\'e dodecahedron \cite{Aurich_Lustig_Steiner_2004c},
and the flat toroidal universe \cite{Aurich_Janzer_Lustig_Steiner_2007}.

\begin{figure}
\begin{center}
\vspace*{-30pt}
\begin{minipage}{11cm}
\hspace*{-20pt}\includegraphics[width=10.0cm]{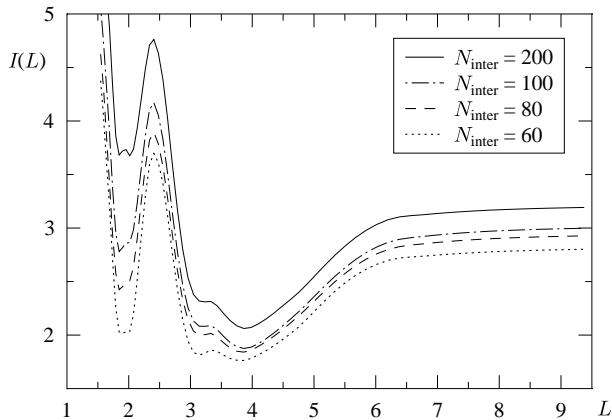}
\end{minipage}
\end{center}
\vspace*{-20pt}
\caption{\label{Fig:I_L_Liu_Li_Mask}
The weighted temperature correlation difference $I(L)$ is computed from
the four correlation functions $C(\vartheta)$
shown in Figure \ref{Fig:C_Theta_Liu_Li_Mask}.
The length $L$ denotes the size of the cubic topology measured
in units of the Hubble length. 
}
\end{figure}

In \cite{Aurich_Janzer_Lustig_Steiner_2007}
we have analysed the WMAP three-year data with respect to the question
whether a flat toroidal universe, i.\,e.\ a cubic topology,
is compatible with the CMB observations.
To that aim the correlation function $C^{\hbox{\scriptsize model}}(\vartheta)$
obtained from the cubic topology with side length $L$
is compared to the observed correlation function
$C^{\hbox{\scriptsize obs}}(\vartheta)$ by using the
integrated weighted temperature correlation difference
\begin{equation}
\label{Eq:I_measure}
I := \int_{-1}^1 d\cos\vartheta \; \;
\frac{(C^{\hbox{\scriptsize model}}(\vartheta)-
C^{\hbox{\scriptsize obs}}(\vartheta))^2}
{\hbox{Var}(C^{\hbox{\scriptsize model}}(\vartheta))}
\hspace{10pt} .
\end{equation}
It is found in \cite{Aurich_Janzer_Lustig_Steiner_2007}
that the results depend on the chosen mask
that is applied to the CMB data.
Applying the kp0 mask to the data shows
that cubic topologies around $L=3.86$ are preferred
whereas using no mask at all leads to models
having a size around $L=4.35$.
($L$ is given in units of the Hubble length
$L_{\hbox{\scriptsize H}} = c/H_0 \simeq 4.26 \hbox{Gpc}$ for $h=0.704$.)
This demonstrates the sensitivity on the pixels
that are selected for the computation of the
correlation function $C(\vartheta)$.

Instead of using the three-year data as in
\cite{Aurich_Janzer_Lustig_Steiner_2007},
we now compute the measure $I(L)$, eq.\,(\ref{Eq:I_measure}),
using the updated five-year data,
where we use the W band foreground reduced map.
Furthermore, we apply the KQ75 mask together with
the four modified Liu-Li masks discussed above.
The result shown in Figure \ref{Fig:I_L_Liu_Li_Mask} is
that cubic topologies slightly smaller than $L=4$
lead to the smallest values of $I(L)$,
i.\,e.\ lead to the best agreement with the data.
The corresponding values for the $\Lambda$CDM concordance model
having infinite volume can also be read off from
Figure \ref{Fig:I_L_Liu_Li_Mask}.
Since the distance to the surface of last scattering (SLS) is
$L_{\hbox{\scriptsize SLS}} = \Delta \eta L_{\hbox{\scriptsize H}}
\simeq 14.2 \hbox{Gpc}$
where $\Delta \eta =\eta_0-\eta_{\hbox{\scriptsize SLS}} = 3.329$
($\eta$ is the conformal time),
topologies above $L\simeq 6.6$ are so large
that the SLS fits completely inside the toroidal universe.
For larger models the value of $I(L)$ saturates at the
infinite limit $L\to\infty$ corresponding to the concordance model.
For all four values of $N_{\hbox{\scriptsize inter}}$
the correlation function $C(\vartheta)$ is in much better agreement
with a cubic universe with $L$ slightly below 4 than
with the infinite $\Lambda$CDM concordance model.
The application of the modified Liu-Li mask has little
influence on the ratio of $I(L)/I(\Lambda\hbox{CDM})$
for the minimum near $L=4$
which is due to the fact that $C(\vartheta)$ is practically
unchanged over the large interval from $\vartheta\simeq 50^\circ$
up to $\vartheta\simeq 165^\circ$ and below $\vartheta\simeq 30^\circ$.
Excluding more pixels,
i.\,e.\ decreasing $N_{\hbox{\scriptsize inter}}$ form 200 to 60,
produces a more pronounced second minimum around $L\simeq 2$
which lies for $N_{\hbox{\scriptsize inter}}=60$ below the
value of the concordance model but above the
absolute minimum near $L=4$.
This new feature is caused by the diminished power of
$C(\vartheta)$ for angles $\vartheta$ in
the interval $[30^\circ,50^\circ]$
leading to a $C(\vartheta)$ with a power suppression already above
$\vartheta \simeq 30^\circ$ instead of the usual $60^\circ$.


\section{Summary}

The presently best full-sky maps of the cosmic microwave background radiation
are obtained by WMAP whose receivers are differential radiometers
measuring the difference between two telescope beams
separated by about $141^\circ$.
Liu and Li \cite{Liu_Li_2008a,Liu_Li_2008b} discuss the possibility
that the inclusion of hot pixels in the map-making process
causes a bias in those pixels corresponding to a scan ring
of about $141^\circ$.
Due to the large diameter of the scan rings,
the affected pixels lie in a region far from the galactic plane
which is usually considered to be scarcely contaminated.
Here we confirm the bias found in \cite{Liu_Li_2008a,Liu_Li_2008b}
and, furthermore,
compute the temperature two-point correlation function $C(\vartheta)$
by taking this bias into account.
The most dramatic effect of these biased pixels is the
so-called correlation hole at $\vartheta=180^\circ$ in $C(\vartheta)$,
which is absent by eliminating sufficiently many affected pixels.
Thus, a correlation function $C(\vartheta)$ is obtained
which displays almost no correlations
above all angular scales larger than $60^\circ$.

In Figures \ref{Fig:Histo_ILC_KQ75}--\ref{Fig:I_L_Liu_Li_Mask},
we have presented our results for the modified Liu-Li mask using the
threshold temperature $T_{\hbox{\scriptsize thres}}=1\,000\,\mu\hbox{K}$
and employing the KQ75 mask.
We have checked that the conclusions of this paper remain the same
if the KQ85 mask is used instead of the KQ75 mask and,
furthermore, if the calculations are performed for the thresholds
$2000\mu\hbox{K}$ and $4000\mu\hbox{K}$, respectively.

Using the correlation function $C(\vartheta)$ based on the most
reliable data,
a test for a universe with a cubic topology is carried out.
It is found that such a model fits $C(\vartheta)$ better than
the $\Lambda$CDM concordance model
if the side length $L$ of the cubic fundamental cell
is slightly smaller than $L=4$ measured in units of the Hubble length.


\section*{Acknowledgments}

HEALPix (healpix.jpl.nasa.gov)
\cite{Gorski_Hivon_Banday_Wandelt_Hansen_Reinecke_Bartelmann_2005}
and the WMAP data from the LAMBDA website (lambda.gsfc.nasa.gov)
were used in this work.
The computations are carried out on the Baden-W\"urttemberg grid (bwGRiD).


\section*{References}

\bibliography{../bib_astro}

\bibliographystyle{h-physrev3}

\end{document}